# The IFF Foundation

# for Ontological Knowledge Organization

## Robert E. Kent[1]

**Abstract:** This paper discusses an axiomatic approach for the integration of ontologies, an approach that extends to first order logic a previous approach (Kent 2000) based on information flow. This axiomatic approach is represented in the Information Flow Framework (IFF), a metalevel framework for organizing the information that appears in digital libraries, distributed databases and ontologies (Kent 2001). The paper argues that the integration of ontologies is the two-step process of alignment and unification. Ontological alignment consists of the sharing of common terminology and semantics through a mediating ontology. Ontological unification, concentrated in a virtual ontology of community connections, is fusion of the alignment diagram of participant community ontologies – the quotient of the sum of the participant portals modulo the ontological alignment structure.

**Keywords:** knowledge organization, ontology, integration, alignment, unification, semantics, classification, model, theory and logic.

## Introduction

The library organizes portions of the bibliography universe (Wynar and Taylor 1992). The bibliographic universe consists of all information-bearing entities (Miksa 1996), whether written, printed, digitized, taped, painted, etc. The library organizes both knowledge and materials. The organization of knowledge and the organization of materials are distinct, but with important interconnections. Knowledge organization underlies most modern library classifications. It involves the registration, evaluation, and classification of "thoughts, ideas, and concepts" in order to adequately represent universal knowledge (Wynar and Taylor 1992).

Many people believe (Miksa 1996) that a new library idea is emerging, a shift from the public space phenomenon of the modern library to a private space phenomenon. New information technologies are transforming the modern library into an emerging library in the same way that earlier libraries were transformed in the 19th century into the modern library. Although knowledge organization has been important in the modern library, it will be even more important in the emerging library. And ontology, a core concept of knowledge organization, will be vitally important in the emerging library. The ontologies of the emerging library will be organized in ontological frameworks. However, the emerging library will eschew a monolithic approach to ontological frameworks, and instead will advocate a modular approach.

The IFF, an ontological framework that supports modularity in a principled fashion, will organize the ontologies of the emerging library by means of a "concept lattice of theories." Because the emerging library is a private space library (Miksa 1996), the integration of ontologies will also be an important concept. Ontological integration uses the paradigm of participant community ontologies formalized as IFF logics, communities interacting through portal logics, a common shared extensible ontology formalized as an IFF theory, participant community alignment links formalizable as IFF theory morphisms, and the free lifting of alignment links from the theoretical context to the logical context. Figure 1 schematically presents the architecture for ontological integration. By using the techniques, principles and terminology of the IFF, this architecture can be constructed through the two-step process (Figure 7) of alignment and unification.

It is the goal of this paper to explain how ontologies and the process of ontological integration are represented with the concepts and terminology of the IFF. This paper extends ontological integration from the theory of information flow (Kent 2000) to first order logic. The paper has seven sections. Sections 2 and 3 provide background information about ontologies and the IFF. Section 4 explains how standard ideas of ontological integration are represented in the IFF. Section 5 presents the two-step process of ontology integration. As a complement to the theoretical approach in the main part of the paper, section 6 gives a related practical approach. Section 7 gives a summary of the study. The Notes section presents the ideas of ontological integration and explains important concepts of the IFF.

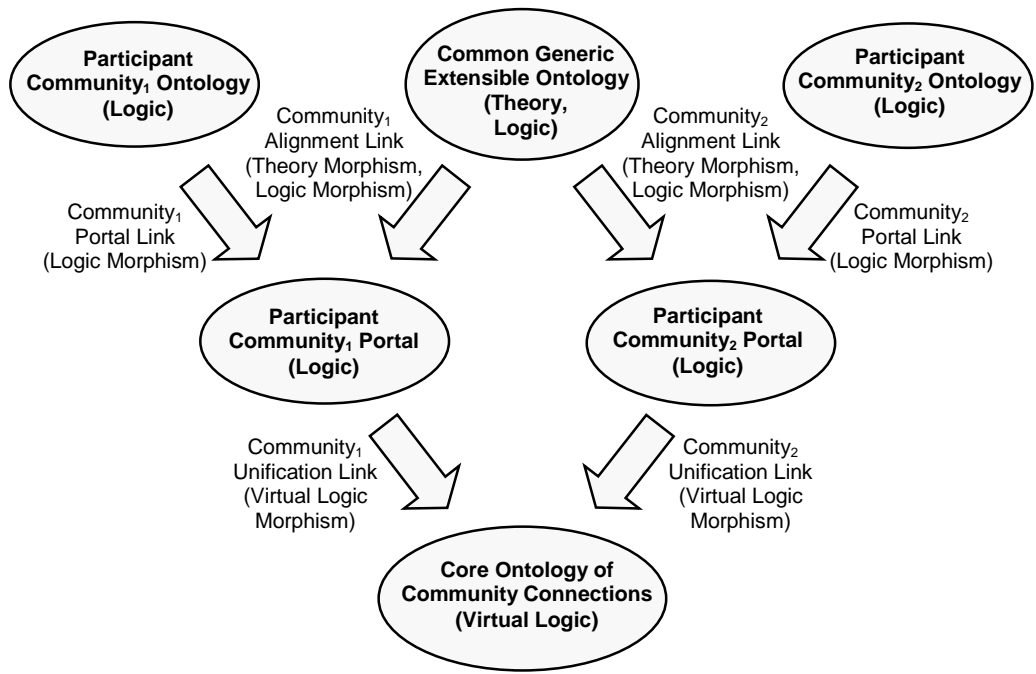

Figure 1: Ontological Integration Process – schema

# Ontologies

Ontology is a branch of metaphysics concerned with the nature and relations of being and the categorical structure of reality. Categories are the most fundamental things that exist or may exist in a domain of discourse. Ontology studies such categories. The word *ontology* comes from the Greek – it is constructed from the prefix *ontos* that means "being" or "existence" and the base *logos* that means "to reason." The expressions in an ontology use a language containing the relevant predications (entity and relation types). Ontological languages are modeled here with the concept of an IFF *type language*[2]. Ontologies and ontology sharing traditionally reside within the field of knowledge representation (Sowa 2000). Knowledge representation applies logic and ontology to knowledge organization. Logic supplies the form, consisting of framework and inferencing capabilities, whereas ontology supplies the content, consisting of the entities, relations, and constraints in the application domain.

The use of ontologies for knowledge organization is both old and new (Kent 2000). In the far past, Aristotle effectively used ontological ideas in his study of predication and his system of classification (his categories). In the near past, Ranganathan effectively used dynamic ontological organizing principles (now understandable as ideas of conceptual and relational knowledge organization) in his development of the Colon faceted classification system for organizing large research libraries. In the emerging library, generic knowledge-organizing frameworks, such as the IFF, advocate the use of ontologies for domains such as e-commerce, bio-informatics, etc. The theory and practice of ontological integration developed in this paper will facilitate interoperability between the knowledge organizations in the emerging library.

Ontologies fall along a formal spectrum. The terminological ontologies or taxonomies, located at the informal end of the spectrum, axiomatize the hierarchy of categories by utilizing the meta-relations of subtyping and disjointness. The WordNet system, developed at Princeton and widely used for natural language processing, is a large example of a terminological ontology. The axiomatized ontologies are located at the formal end of the spectrum. The Cyc project, whose avowed purpose is to model common-sense knowledge and reasoning, is a large example of an axiomatized ontology.

One of the earliest ontologies created was described in the *Categories* of Aristotle (Aristotle 350 B.C.E.). Aristotle's ontology was a simple upper-level ontology involving four major categories: substance (the primary category), quality, quantity and relation; in addition, there were several minor subcategories of relation: determination in time and space, position, condition, action and passion or passivity.

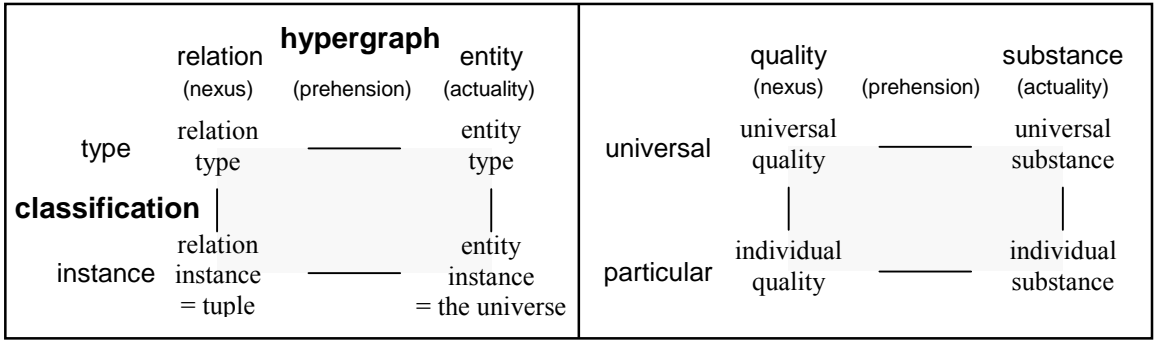

**Figure 2a: The Information Flow Framework – mathematical knowledge organization**

**Figure 2b: Aristotle's ontological framework – philosophical knowledge organization**

However, a more important development implicit in Aristotle's *Categories* is an ontological framework based upon his theory of predication. For Aristotle a statement (proposition) consists of a subject, which is what the statement is about, and a predicate, which is what the statement says about its subject. For Aristotle there are two kinds of predication: essential predication and accidental predication.

○ An essential predication is something that is said-of a subject. This is the relation of ontological classification between a thing and a natural kind. Essential predication is represented with the concept of an IFF *classification* (Kent 2000).

○ An accidental predication is something that is being-in a subject. This is the relation of ontological dependence between a substance and a quality. Accidental predication is represented with the concept of an IFF *hypergraph*[3].

Essential predication divides things into universals (types) and particulars (instances). Accidental predication divides things into substances (entities) and qualities (relations). The mathematical knowledge organization used in this paper is represented by the semantic architecture of the IFF (Figure 2a) and concentrated in the concept of an IFF *model*[4]. An IFF model represents "interpretative semantics." This semantic architecture is comparable to the structure of Aristotle's ontological framework (Figure 2b). The two distinctions in Aristotle's ontological framework of "universal versus particular" and of "quality versus substance" are analogous to the two distinctions in the semantic architecture of the IFF between "type versus instance" and "relation versus entity."

# The Information Flow Framework

The Information Flow Framework (IFF) is designed to realize the goal of interoperability between distributed software applications, database applications and knowledge bases. It provides an approach for representing distributed logic, ontologies, and knowledge representation. The IFF is a metalogic that uses modularity to support semantic interoperability among object-level ontologies. The IFF takes a building blocks approach towards the development and utilization of object-level ontological structure. This rather elaborate categorical approach uses insights from the theory of information flow (Barwise and Seligman 1997) and the theory of formal concept analysis (Ganter and Wille 1999). In this paper, we show how the IFF provides a foundation for the integration of ontologies in a distributed setting. In turn, the idea of ontological integration illustrates the intuitions behind the IFF foundation for ontological knowledge organization.

The IFF provides a principled foundation for sharing, manipulating, relating, partitioning, composing, and discussing ontologies. The IFF represents metalogic, and as such operates at the structural level of ontologies. In the IFF, there is a precise boundary between the metalevel and the object level. The terminology and axiomatics of the IFF is located in various meta-ontologies. The IFF is partitioned into three metalevels: top, upper and lower. These metalevels correspond to the set-theoretic distinction in foundations between the generic, the large and the small.

○ The top metalevel of the IFF is used to represent and axiomatize the upper metalevel. It does this by providing an interface between the KIF logical language and the upper metalevel of the IFF. The top metalevel contains the basic KIF meta-ontology.

○ The upper metalevel of the IFF is used to represent and axiomatize the lower metalevel. It does this by representing category theory, information flow and formal concept analysis. The upper metalevel contains the core, classification and category theory meta-ontologies.

○ The lower metalevel of the IFF is used to represent and axiomatize the object metalevel. It represents the structural aspect of the Standard Upper Ontology (SUO) using a categorical expression. The lower metalevel contains the core, classification, model theory and algebraic theory meta-ontologies. The IFF model theory ontology (IFF-MT) is the central ontology in the lower metalevel. It represents a somewhat novel version of model theory, whereas the closely related algebraic theory ontology represents universal algebra. In support of the IFF-MT are the lower classification and lower core ontologies.

In order to formulate ontological integration in terms of the IFF this paper uses the terminology and semantics of the IFF-MT. Model theory is the study of the interpretations of any language, formal or natural, by means of set-theoretic or category-theoretic structures. Apart from the use of set theory or category theory, model theory is agnostic about the kinds of things that exist. The latter are described and axiomatized in object-level ontologies. In this paper, we represent an ontology as either an IFF *theory*[5] or an IFF *logic*[6]. An ontology $O$ is represented as a theory $T$ if it has no instances. A theory represents formal semantics. Otherwise, an ontology $O$ is represented as a logic $L$. A logic represents a combined semantics, both a formal semantics and an interpretative semantics.

## The Nature of Integration[7]

In this section, we formulate various notions of ontological integration in terms of the IFF. In the following, reference is made to the endnotes where John Sowa's glossary definitions (Sowa 2000, Appendix B.1) for ontological integration are paraphrased with respect to two participant community ontologies (Figure 3). These glossary definitions define the terminology of methods and techniques for defining, sharing, and merging ontologies. According to Sowa, this terminology and these notions originated from a working group of the International Committee for the Information Technology Standards (NCITS).

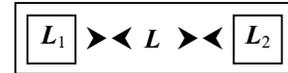

**Figure 3: Integration**

This section represents these notions with concepts from the IFF: refinement is represented as appropriate morphisms in the theoretical or logical context; alignment is represented by the creation of a span of morphisms over representations for the community ontologies; and aligned unification is represented as a pushout of the alignment span.

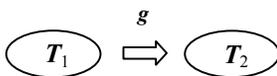

**Figure 4a: Refinement – abstract**

| entity type | $\alpha_1 \mapsto \alpha_2$ | entity type |
|---|---|---|
| function type | $\phi_1 \mapsto \tau_2$ | term |
| relation type | $\rho_1 \mapsto \phi_2$ | expression |

**Figure 4b: Refinement – details**

First, we represent the refinement operator. This representation is key, since alignment and unification are expressed in terms of refinement. A primitive type is represented as an IFF type – an entity type, a function type or a relation type. A composite type is represented as an IFF term or an IFF expression. A composite type generalizes a primitive type – terms generalize function types and expressions generalize relation types. Entity types are not composite. The general notion of refinement[8] maps the entity types of the first ontology to entity types of the second ontology, but maps the function or relation types of the first ontology to terms and expressions of the second ontology, respectively (Figure 4b). From this original formal viewpoint, an ontology is represented as an IFF theory, a refinement is represented as an IFF theory interpretation (Figure 4a), and the partial ordering of refinements is represented as composition of IFF theory interpretations. From an extended composite viewpoint, an ontology is represented as an IFF logic, a refinement is represented as an IFF logic interpretation, and the partial ordering of refinements is represented as composition of IFF logic interpretations.

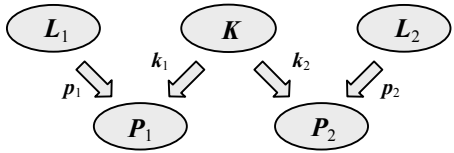

| equivalent function types | linked function types |
|---|---|
| $\phi_1 \cong \phi_2$ | $\phi_1 \leftharpoondown \phi \rightharpoonup \phi_2$ |
| equivalent relation types | linked relation types |
| $\rho_1 \cong \rho_2$ | $\rho_1 \leftharpoondown \rho \rightharpoonup \rho_2$ |

**Figure 5a: Alignment Diagram – abstract**

**Figure 5b: Alignment or Partial Compatibility – details**

Next, we represent alignment and partial compatibility[9]. The intent of alignment is that mapped categories are equivalent. To appropriately formalize this, we represent an equivalence pair of types as a single type in a mediating ontology, with two mappings from this new type back to the participant community types (Figure 5b). This implies that alignment and partial compatibility can be represent as a kernel span of theory morphisms (Figure 5a). The mediating ontology $K$ represents both the equivalenced categories and the axiomatization needed for the degree of compatibility, partial or complete. Since the theoretical alignment links $k_1$ and $k_2$ preserve this axiomatization, compatibility will be enforced. However, new subtypes or supertypes may need to be introduced in extension ontologies called portals in order to provide suitable targets for alignment. This results in a 'W'-shaped diagram of logic morphisms (Figure 5a) with the logical portal links $p_1$ and $p_2$ connecting the participant community ontologies to their portal logics $P_1$ and $P_2$. The direction of the portal links is compatible with the unification diagram below.

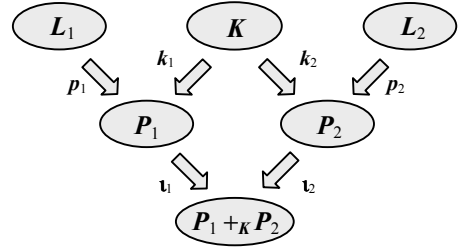

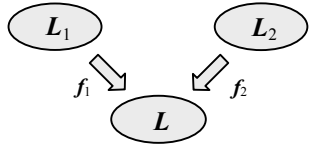

**Figure 6a: Initial (nonaligned) Unification Diagram**

**Figure 6b: Aligned Unification Diagram**

Finally, we represent unification[10]. In the usual approach to unification, the refinements are represented by theory/logic morphisms from the two participant community ontologies to refined ontologies, where the latter are isomorphic. Because of this isomorphism, we can replace these refinements with a single theory/logic $L$. We end up with an opspan of theory/logic morphisms (Figure 6a) consisting of two morphisms $f_1$ and $f_2$ with a common target theory/logic. However, although we have a representation in terms of the IFF, we feel that this opspan representation is too loose, since it is not aligned. In order to tighten this up we assume that the resulting opspan comes from taking the fusion of an alignment span of logic morphisms (Figure 6b).

## The Process of Integration

Integration is a two-step process of (1) alignment and (2) unification with respect to that alignment. Unification is the automatic process of fusion in the theoretical/logical context. However, alignment is not automatic. We start (Figure 7) with two participant community ontologies represented as logics $L_1$ and $L_2$. We end, after alignment and unification, with an opspan of logic morphisms

$$f_1 : L_1 \rightleftharpoons L \text{ and } f_2 : L_2 \rightleftharpoons L$$

that represents integration of the participant community ontologies through a mediating ontology $K$. The following discusses the process.

## *Alignment*

Alignment is the process of specifying a diagram (Figure 7) in the theoretical/logical context that connects the participant community ontologies through common agreement. This alignment diagram is used for unification. Four questions need to be asked and answered for ontological alignment: where, what, who and a duplicate how – a how for where and a how for what.

**Where** do we want to interact with the other community and **how** is this place of interaction related to our community ontology? Where is the locus of integration? The question of "where" refers to the local *portal*, the logic we use for interaction. The question of "how" refers to the *portal link* for our community. Each community ontology involved in integration needs to localize its activities vis-à-vis integration. This locus of integration is called a portal – a portal is a communicating part or area of a system. The portals are represented by logics $P_1$ and $P_2$ and the portal links are represented by logic morphisms

$$p_1 : L_1 \rightleftharpoons P_1 \text{ and } p_2 : L_2 \rightleftharpoons P_2$$

that connect the participants with their locus of integration (Figure 7). One reason for a portal is the incorporation into community ontologies of new subtypes or supertypes necessary for alignment. This incorporation extends the type aspect of the participant community ontology.

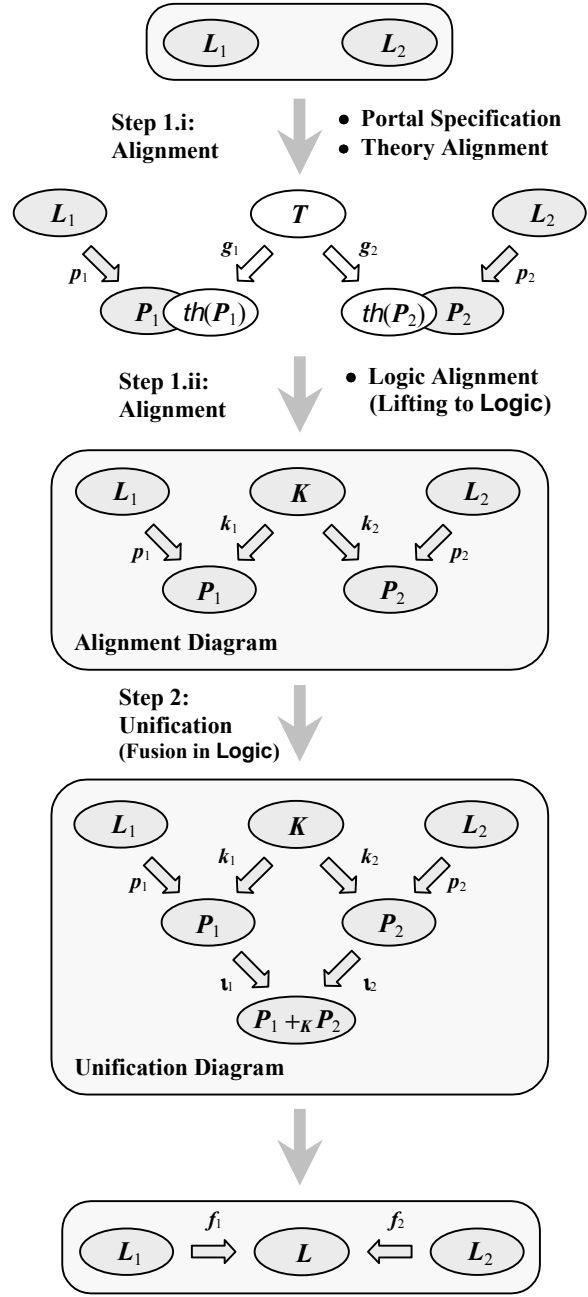

**Figure 7: Ontological Integration Process – details**

**What** do we want to say? What is the common semantics? What common meaning do we want to express? What is the expression for the *mediating ontology*? What is the language and theory of the mediating logic? The question about "what" refers to the mediating ontology – what is the language and theory of the mediating ontology? **How** do we say it in our own terms? How is this common semantics expressed in our own world? How does our community formalize the common semantics? The "how" question refers to how we express the *alignment link* for our community; that is, how do we specify the theory interpretation from the theory of the mediating logic to the theory of our community logic? To answer the "what" question, we assume that the participant ontologies $L_1$ and $L_2$ agree to specify their interaction via a theory $T$. This will be the underlying theory $T = th(K)$ of the mediating ontology $K$. The types of $T$ (entity, function and relation) represent the aligned categories. The axioms of $T$ represent the amount and nature of the desired compatibility. To answer the "how" question, each community ontology specifies its own theoretical alignment link. These are represented as theory morphisms

$$g_1 : T \Rightarrow th(P_1) \text{ and } g_2 : T \Rightarrow th(P_2)$$

that connect the mediating theory to the underlying theories of the participant portals.

**Who** do we want to talk about? What universe of discourse can serve as a base for the common semantics? What is the universe for the mediating logic? The answer to the question of "who" do we want to talk about has two parts: (1) use the free logic[11] $K = log(T)$ over the mediating theory $T = th(K)$, and (2) transform by adjointness the theoretical alignment links

$$g_1 : T \Rightarrow th(P_1) \text{ and } g_2 : T \Rightarrow th(P_2)$$

into the equivalent logical alignment links

$$k_1 : log(T) \rightleftharpoons P_1 \text{ and } k_2 : log(T) \rightleftharpoons P_2$$

that connect the mediating logic to the participant portals. Logical alignment lifts the theoretical alignment link to the free logic $L = log(T)$. The logical alignment links are the composition of the application of the logic operator to the theoretical alignment links

$$log(g_1) : log(T) \rightleftharpoons log(th(P_1)) \text{ and } log(g_2) : log(T) \rightleftharpoons log(th(P_2)),$$

followed by the canonical logic morphisms (counit components)

$$\varepsilon_{L1} : log(th(P_1)) \rightleftharpoons P_1 \text{ and } \varepsilon_{L2} : log(th(P_2)) \rightleftharpoons P_2,$$

which are the identity on types and the intent morphisms on instances (entity and relation). The use of free logics has been successfully used in the context of the theories and logics of information flow (Kent 2000). This paper extends that technique to free first order logics[12].

## *Unification*

The process of unification is fusion of the alignment diagram. We use the logic alignment diagram to specify a logic invariant: we use the type aspect to specify an equivalence relation on the types (entity, function and relation) of the sum logic, and we use the instance aspect to specify an appropriate subset of the instances (entity, function and relation) of the sum[13] logic $P_1+P_2$. This logic invariant[14] induces a quotient logic[15] $P_1 +_K P_2$ over the sum logic: types are the quotient classes of the corresponding equivalence relation, whereas instances are the members of the instance subsets of the logic invariant. This virtual ontology is a fusion[16] of the community portals with respect to the alignment diagram. It represents the complete system of ontological integration. A canonical logic morphism

$$q : P_1 + P_2 \rightleftharpoons P_1 +_K P_2$$

links the sum logic to the quotient logic – its type component maps sum types to their equivalence class, and its instance component is subset inclusion. The fusion injections

$$\tilde{\imath}_1 = \imath_1 \cdot q : P_1 \rightleftharpoons P_1 + P_2 \rightleftharpoons P_1 +_K P_2 \text{ and } \tilde{\imath}_2 = \imath_2 \cdot q : P_2 \rightleftharpoons P_1 + P_2 \rightleftharpoons P_1 +_K P_2.$$

are the composition of the sum injections $\imath_1$ and $\imath_2$ with the canonical logic morphism $q$. In the Figure 7, the final integration opspan is define as by the compositions

$$f_1 = p_1 \cdot \tilde{\imath}_1 \text{ and } f_2 = p_2 \cdot \tilde{\imath}_2.$$

There are two extremes of ontological integration – nothing or everything. Any two logics can be vacuously integrated by forming the binary sum logic. This has no (empty) alignment and hence represents the trivial integration. Any logic can be completely integrated with a copy of itself. This has full (identity) alignment and complete integration.

# A Practical Alternative

The main part of this paper has discussed a theoretical approach to ontological integration. However, a practical approach also may be of interest. In both the theoretical and practical approaches, the participant community ontologies need to specify and interpret a mediating theory for integration. Beyond this, the two approaches differ. To answer the question of "who" in the practical approach, we assume that the logics $L_1$ and $L_2$ of the two community ontologies agree to restrict their attention to a common part of both worlds called $C$. This means that the universes of the two community ontologies intersect and that $C \subseteq univ(L_1) \cap univ(L_2)$ is a subset of this intersection. Define the portals $P_1 = L_1@C$ and $P_2 = L_2@C$ to be the restriction of the full community logics to the sub-universe $C$, and define the portal links

$$p_1 : L_1 \Rightarrow L_1@C \text{ and } p_1 : L_1 \Rightarrow L_1@C$$

to be the restriction logic morphisms which are the identity on types and the inclusion on instances. The practical approach must answer the additional question: **Can** communities agree? Can all communities agree upon the meaning of the theory of the mediating ontology? The interpretative semantics that a participant community ontology gives to the mediating theory is specified by its theoretical alignment link. Do the interpretative semantics for all communities agree? In the practical approach, to answer the question of "can," the logic fiber operators $g_1^{-1}$ and $g_2^{-1}$ of the theoretical alignment links are applied to the portal logics $L_1@C$ and $L_2@C$, respectively. In order for alignment to succeed, the two logics that result must be the same: $g_1^{-1}(L_1@C) = L@C = g_2^{-1}(L_2@C)$. This is defined to be the logic $K = L@C$ of the mediating ontology. This is not the free logic $log(T)$, but there is a logic morphism

$$log(T) \rightleftharpoons L@C$$

that connects the free logic to this mediating logic: it is the identity on types and maps an instance in $C$, whether entity or relation, to its intent. Then the fusion logic $L_1{}^{@}C +_{L@C} L_2{}^{@}C$ has the fusion theory $th(L_1) +_T th(L_2)$ as its underlying theory and has $C$ as its universe. There is a logic morphism

$$(L_1{}^{@}C +_{log(T)} L_2{}^{@}C) \rightleftharpoons (L_1{}^{@}C +_{L@C} L_2{}^{@}C)$$

from the free fusion to the fusion at *C*: it is the identity on types and the diagonal function on instances in *C*.

In summary, in the practical approach to alignment there are three agreements between communities: the universe of the mediating ontology is the first agreement: *univ*(*K*) = *C*; the theory of the mediating ontology is the second agreement: *th*(*K*) = *T*; and the identification of the mediating ontology with the fiber image of the portals along the theoretical alignment links is the third agreement.

## Summary

The modern library is being transformed, and a new library is emerging (Miksa 1996). Ontologies and the integration of ontologies in a distributed environment (Sowa 2000, Appendix B.1) will be important and necessary concepts in the knowledge organization of the emerging library. This paper has discussed how the modular approach for ontologies and ontological integration, which is needed in the emerging library, can be represented in the metalogic of the Information Flow Framework (Kent 2001). This first order logic and model-theoretic representation is an extension of the information flow representation for ontology sharing (Kent 2000).

## Notes

[1] Robert E. Kent has a bachelors degree in mathematics from Southern Illinois University, a masters degree in mathematics from the University of California at Santa Barbara, and a doctorate degree in engineering from the University of California at Los Angeles. He has previously designed two markup languages for distributed knowledge representation: the Ontology Markup Language (OML) and the Conceptual Knowledge Markup Language (CKML). Robert is a voting member of the Standard Upper Ontology working group, where he has authored the starter documents for the Information Flow Framework. He currently resides in Pullman, WA, USA.

[2] It has been the opinion of many that the best way to handle the multivalent relations in ontologies is with hypergraphs. A first order IFF *type language* is a kind of aligned hypergraph. It consists of sets for variables, entity types corresponding to hypergraph nodes, and relation types corresponding to hyperedges, and functions for defining the reference, signature and arity of relation types. In contrast to the notion of a hypergraph, a type language is aligned along its reference function. The set of entity types linked by a relation type is called its signature. Type languages are related through type language morphisms. A type language morphism from source type language to target type language maps source entity (relation) types to target entity (relation) types, preserving signature and arity.

Any type language can be extended to a type language of expressions, which has the same sets for variables and entity types and the same reference function, but has expressions as its relation types. The set of expressions is defined recursively and is the disjoint union of atomic expression, negation conjunction, dis-junction, implication, existential and universal quantifications, and substitutions. There is an embedding type language morphism from any type language to its expression type language.

[3] A hypergraph is similar to a graph but allows edges that link more than two nodes. The set of nodes linked by a hyperedge is called its tuple. An IFF *hypergraph* consists of sets for names, nodes, and hyperedges, and functions for defining the tuple and arity of hyperedges. Hypergraphs are related through hypergraph morphisms. A hypergraph morphism from source hypergraph to target hypergraph maps source nodes (hyperedges) to target nodes (hyperedges), preserving tuple and arity.

[4] The IFF gives a (somewhat novel) category-theoretic axiomatization for first-order model theory based upon the two fundamental ideas of classification and hypergraph (see the two dimensional structure in Figure 2a formed out of classifications along one dimension and hypergraphs along the other dimension). In one sense, an IFF *model* is a hypergraph of classifications. In place of nodes, there is a classification of entities, and in place of hyperedges, there is a classification of relations. In another sense, an IFF model is a classification of hypergraphs – the instance aspect of a model forms an instance hypergraph, and dually the type aspect of a model forms a type language. IFF models are equivalent to the models of traditional many-sorted logic. In this equivalence, the extent functions of the entity, relation and expression classifications are regarded as interpretation functions.

The IFF has a lax notion of satisfaction for tuples. For a tuple to satisfy an expression, or that expression to hold for the tuple, we only require that the arity of the expression be a subset of the arity of the tuple and that the restriction of the tuple to that subset satisfy the expression in the usual sense. There is an expression classification, where expressions classify tuples. When a tuple is classified by an expression, we say that the expression holds for that tuple. A model satisfies an expression in its type language when that expression holds for all tuples; i.e., has maximal intent. In other words, a model for an expression is a model that satisfies the expression. Satisfaction is defined recursively. Models are related through model morphisms. A model morphism is an infomorphism along the classification dimension and a hypergraph morphisms along the hypergraph dimension. Models and model morphisms form the Model context.

[5] An IFF *theory* is a pair consisting of an underlying type language and a set of expressions of that language called axioms. A model for a theory is a model that satisfies every axiom of the theory. A theory logically entails an expression of its underlying language when every model of the theory is also a model of the expression. Obviously, all axioms are entailed by a theory. The closure of a theory is the theory whose axioms consist of all expressions entailed by the original theory. An axiom of the closure is called a theorem of the original theory. Any axiom is a theorem. A theory is closed when it is its own closure. Every model generates a theory, whose underlying language is the type language of the model, and whose axioms are all expressions satisfied by the model. The theory of a model is closed. Theories are related through theory interpretations. A theory morphism from source theory to target theory is a language morphism that maps source axioms to target theorems. Theories and theory morphisms form the Theory context.

[6] An IFF *logic* is an inclusive idea combining the notions of model and theory into a (not necessarily sound) whole. It consists of a theory of types and a model of instances that share a common type language. The theory provides the formal semantics and the model provides the interpretative semantics. In addition, two special subsets are highlighted: there is a subset of the universe called the normal entities and a subset of the tuples called normal tuples, where the components of normal tuples are normal entities and normal entities and normal tuples satisfy the axioms of the theory. A logic is *sound* when every instance of the universe and every tuple is normal. For any logic, the sound part of the logic is obtained by throwing away all abnormal instances and restricting the entity and relation classifications to normal instances. In this paper, we limit ourselves to sound logics, since these enable ontology integration. A logic morphism from source logic to target logic is a theory morphism along the theory aspect and a model morphism along the model aspect. Logics and logic morphisms form the Logic context.

[7] *Integration* is the process of finding commonalities between two community ontologies $O_1$ and $O_2$ and deriving a new ontology $O$ that facilitates interoperability between computational systems that are based on the $O_1$ and $O_2$ ontologies. The new ontology $O$ may replace $O_1$ or $O_2$, or it may be used only as an intermediary between a system based on $O_1$ and a system based on $O_2$. Depending on the amount of change necessary to derive $O$ from $O_1$ and $O_2$, different levels of integration can be distinguished: alignment, partial compatibility, and unification.

[8] One ontology $O_2$ is a *refinement* of another ontology $O_1$ when there is an alignment of every type of $O_1$ to some type of $O_2$. Every type in $O_1$ must correspond to an equivalent type in $O_2$, but some primitive types (functions or relations) of $O_1$ might be equivalent to nonprimitives types (terms or expressions) in $O_2$. Refinement defines a partial ordering of ontologies. If $O_2$ is a refinement of $O_1$, and $O_3$ is a refinement of $O_2$, then $O_3$ is a refinement of $O_1$; if two ontologies are refinements of each other, then they must be isomorphic.

[9] An *alignment* of two ontologies $O_1$ and $O_2$ is a mapping of types (entity, function and/or relation) between two ontologies $O_1$ and $O_2$ that preserves the partial ordering by subtypes in both $O_1$ and $O_2$. If an alignment maps an entity $\varepsilon_1$, function $\phi_1$ or relation $\rho_1$ in ontology $O_1$ to an entity $\varepsilon_2$, function $\phi_2$ or relation $\rho_2$ in ontology $O_2$, then $\varepsilon_1$ and $\varepsilon_2$ are said to be equivalent, $\phi_1$ and $\phi_2$ are said to be equivalent, and $\rho_1$ and $\rho_2$ are said to be equivalent. The mapping may be partial – many types (entities, function or relations) in $O_1$ or $O_2$ may have no equivalents in the other ontology. Before two ontologies $O_1$ and $O_2$ can be aligned, it may be necessary to introduce new subtypes or supertypes of concepts or relations in either $O_1$ or $O_2$ in order to provide suitable targets for alignment. No other changes to the axioms, definitions, proofs, or computations in either $O_1$ or $O_2$ are made during the process of alignment. A *partial compatibility* of two ontologies $O_1$ and $O_2$ is an alignment that supports equivalent inferences and computations on all equivalent types. If $O_1$ and

$O_2$ are partially compatible, then any inference or computation, that is expressible in one ontology using only aligned types, can be translated to an equivalent inference or computation in the other ontology.

[10] *Unification* or total compatibility may require extensive changes or major reorganizations of $O_1$ and $O_2$, but it can result in the most complete interoperability: everything that can be done with one can be done in an exactly equivalent way with the other. A unification is a one-to-one alignment of all entities and relations in two ontologies $O_1$ and $O_2$ that allows any inference or computation expressed in one to be mapped to an equivalent inference or computation in the other. The usual way of unifying two ontologies $O_1$ and $O_2$ is to refine each to more refined ontologies $\tilde{O}_1$ and $\tilde{O}_2$ whose categories are one-to-one equivalent.

[11] For any theory $T$, the *free logic* $log(T)$ is defined as follows. Its underlying theory is $T$. Its underlying model is the free model over $T$, whose entity classification is the power classification over the entity types of $T$ and whose relation classification is a modification of the power classification over the relation types of $T$. Instead of the entire power set of relation types, the tuples in $log(T)$ consist of the subsets of relation types whose arity is a subset of a particular indexing arity. This is a kind of power classification over the arity function of $T$. The instance arity function maps such a subset of relation types to their indexing arity. The instance signature function maps such a subset of relation types to the signature with indexing arity whose coordinate at any variable in that arity is the set of same coordinates for the signatures of all relation types whose arity contains that variable; that is, for any subset of variables $X$ and any arity fiber subset $R$ of relation types and any variable $x \in X$, define the instance signature function $\partial_0(R)_x = \{\partial_L(\rho)_x \mid \rho \in R, x \in arity(L)(\rho)\}$.

[12] A *free logic* $log(T)$ over a theory $T$ has the following meaning: any theory morphism $g$ from a theory $T$ to the underlying theory $th(L)$ is equivalent to a logic morphism $\hat{g}$ from the free logic $log(T)$ to the logic $L$. This equivalence can be computed as the composition $\hat{g} = log(g) \cdot \varepsilon_L$, where

$$\varepsilon_L : log(th(L)) \rightleftharpoons L,$$

the $L^{\text{th}}$ component of the counit, is the identity on types and maps entity instances to their entity extents and maps relation instances (tuples) to their relation extents consisting of all relation types (with possibly smaller arities) that classify that tuple. In Information Flow, the notion of power logic was used in the first step (**Lifting** to Logic) of the two-step process for ontology sharing (Kent 2000). That **Lifting** to Logic step corresponds to the logical alignment substep (Figure 7) of first order logics.

When free logics are investigated with respect to the lax IFF notion of satisfaction, there is a question about the relationship between unary relation types (unary predicates) and entity types (sorts). The IFF adopts a laissez-faire attitude: use the least constraints needed. It defines its basic logic using the constraint that every unary predicate have a particular sort (the single sort in its signature): the relational extent of the unary predicate is contained in the entity extent of its sort. However, the free logic construction seems to require some tightening up. The logic notion needed for the free logic construction further requires that every sort be the sort of some unary predicate that is extensionally equivalent to it.

[13] Given two models $A_0$ and $A_1$, the *sum* $A_0+A_1$ is the model, whose instance hypergraph is the product of the component instance hypergraphs, whose type language is the sum of the component type languages, and whose entity (relation) classification is the sum of the component entity (relation) classifications. Each sum model comes equipped with two injection model morphisms from component models to sum. These are defined to be the sum injection on types and the product projection on instances. The sum model is a binary coproduct in Model, the model-theoretic context. Given two theories $T_0$ and $T_1$, the *sum* $T_0+T_1$ is the theory, whose type language is the sum of the component languages, and whose axiom set is the union of the component axiom sets. The two underlying language sum injection morphisms are theory sum injection morphisms. The theory sum is a coproduct in Theory, the theoretical context. Given two (sound) logics $L_0$ and $L_1$, the *sum* $L_0+L_1$ is the logic, whose model is the sum of the underlying models, and whose theory is the sum of the underlying theories. The underlying model sum injections are logic sum injections – the type functions are also theory morphisms. The logic sum is a coproduct in Logic, the logical context.

[14] For any model $A$, a *dual invariant* is a pair $J = (inst(J), typ(J))$ consisting of a sub-hypergraph $inst(J)$ of the instance hypergraph $inst(A)$ and an endorelation $typ(J)$ on the type language $typ(A)$. For any theory $T$, an *theory endorelation* $J$ is a type language endorelation on its underling type language. For any (sound) logic $L$, a *dual invariant* is a pair $J = (mod(J), th(J))$ consists of a dual invariant $mod(J)$ on the underlying model $mod(L)$ and an endorelation $th(J)$ on the underlying theory $th(L)$, where the type language endorelation of the model dual invariant is the same as the type language endorelation of the theory endorelation: $typ(mod(J)) = lang(th(J))$.

[15] The *dual quotient* of $A$ by a model dual invariant $J$ is the model $A/J$, whose instance hypergraph is $inst(J)$, whose type language is the quotient $typ(A)/typ(J)$, whose entity classification is the quotient classification $ent(A)/ent(J)$, and whose relation classification is the quotient classification $rel(A)/rel(J)$. The *quotient* of $T$ by a type language endorelation $J$, written $T/J$, is the theory whose type language is the quotient type language $lang(T)/lang(J)$, and whose axiom set consists of all "equivalence images" $[\varphi]_{lang(J)}$ for all $\varphi \in axm(T)$. The *dual quotient* of $L$ by a logic dual invariant $J$, written $L/J$, is the logic whose model is $mod(L)/mod(J)$ and whose theory is $th(L)/th(J)$. This is well-defined, since any relation instance $t$ in $inst(mod(L))$ satisfies any constraint $[\varphi]_{th(J)} \in axm(th(L)/th(J)))$.

[16] For any two logic infomorphisms $f_0 : L \rightleftharpoons L_0$ and $f_1 : L \rightleftharpoons L_1$ that share a common source logic $L$, we define a dual invariant $J = (A, R)$ on the sum $L_0+L_1$. Let $A$ be the pairs of instances on which the instance functions of $f_0$ and $f_1$ agree, and let $R$ be the binary relation between types of $L_0$ and $L_1$ that are linked by a type of $L$ through the type functions of $f_0$ and $f_1$. The *fusion* logic is the dual quotient $L_0+_L L_1 = (L_0+L_1)/J$. With the injection logic morphisms, the fusion logic is a pushout in Logic, the logical context.